\documentclass[
    ,final            % use final for the camera ready runs
    ,numberedheadings % uncomment this option for numbered sections
  ]
  {aipproc}

\layoutstyle{6x9}

\begin{document}

\title{Scalar Mesons in Charm Decays}

\classification{13.25.Ft,14.40.Lb,14.40.Cs,11.80.Et}
\keywords      {Charmed mesons; scalar mesons; Dalitz plots}

\author{Jonathan L. Rosner}{
  address={Enrico Fermi Institute, University of Chicago,
Chicago, IL 60637}
}
\begin{abstract}
Results on light scalar mesons in charmed particle decays studied by the CLEO
Collaboration at the Cornell Electron Storage Ring are reviewed.
\end{abstract}

\maketitle

\section{Introduction}

The CLEO Collaboration, working at the Cornell Electron Storage Ring (CESR),
has performed studies of charmed meson and charmonium decays, whose three-body
final states will yield rich information on low-mass scalar resonances once
fully analyzed.  We have observed the $f_0(980)$ and $a_0(980)$, broad S-wave
amplitudes ``$\sigma(600)$'' and ``$\kappa(800)$'' in $\pi \pi$ and $K \pi$,
and evidence for $f_0(1370) \to \pi \pi$.  Data sets include open charm
production near 10 GeV, 27.5 million $\psi(2S)$ as a source of $\chi_c$ states
and about 6 million tagged $\pi^+ \pi^- J/\psi$, $818 \pm 8 {\rm ~pb}^{-1}$ at
$\psi(3770)$ leading to more than 5 million very clean $D \bar D$ pairs, and
about 600 pb$^{-1}$ at 4170 MeV, yielding about 570 thousand $D_s \bar D_s$
pairs.  Production mechanisms can affect the determination of resonance
parameters
(especially for broad states).  In this report we give some examples of CLEO's
results on scalar mesons obtained from charmed meson and charmonium decays.

\section{Examples of channels}

CLEO has data on a wide variety of three-body charmed meson and charmonium
decays.  These include the following:

\begin{itemize}

\item $D^+ \to K^- \pi^+ \pi^+$ \cite{:2007nn}: Is there a $\kappa$ in the
low-energy $K \pi$ S wave?

\item $D^0 \to K_S \pi^+ \pi^-$: CLEO's sample of 9 fb$^{-1}$ near 10 GeV
needed no $\sigma$, $\kappa$ \cite{Muramatsu:2002jp} but BaBar
\cite{Pappagallo:2007zz} and Belle \cite{Poluektov:2006ia} have $m(\sigma_1)
\simeq 500$ MeV, $m(\sigma_2) \simeq 1037$ MeV.

\item $D^0 \to \pi^0 \pi^+ \pi^-$ is dominated by $\rho^\pm,\rho^0$
bands \cite{CroninHennessy:2005sy}.

\item $D^+ \to \pi^- \pi^+ \pi^+$:  scalars appear to be
important \cite{Bonvicini:2007tc}.

\item $D^0 \to K^+ K^- \pi^0$: A $K \pi$ S-wave amplitude is needed
\cite{Cawlfield:2006hm}.

\item $D^0 \to K_S \eta \pi^0$:  $a_0(980)$, $K^*(892)$ are seen
\cite{Rubin:2004cq}.

\item $D^+ \to K^-K^+\pi^+$:  a $K^- \pi^+$ S-wave (e.g., the LASS amplitude
\cite{LASS}) is important.

\item $D^0 \to K_S \pi^0 \pi^0$: Subsystems include $K^*(892)$,
$f_0(980)$, $f_0(1370)$, $K^*(1680)$ \cite{Naik:2007es}.

\item $\chi_{c1} \to \eta \pi^+ \pi^-$ involves $a_0(980)$, $f_2(1270)$, and
$\sigma(500)$; $\chi_{c1} \to (K^+ K^- \pi^0,~\pi^\pm K^\mp
K_S)$ involves $K^*(892)$, $K_0^*(1430)$, $K_2^*(1430)$, and $a_0(980)$
\cite{Athar:2006gh}.

\end{itemize}

\section{$D^+ \to K^- \pi^+ \pi^+$}

$D^+$ candidates for $K^- \pi^+ \pi^+$ are selected on the basis of energy
and momentum conservation.  The sample based on 572 pb$^{-1}$ (about 2/3 of
the final total) \cite{Bonvicini:2008jw}, superseding an earlier one
\cite{:2007nn} based on 281 pb$^{-1}$, contains 140793 events with a small
background of 1.1\%.  The largest previous sample was $\sim 15,000$ events
from Fermilab E791 \cite{E791}.

\begin{figure}
\mbox{\includegraphics[width=0.47\textwidth]{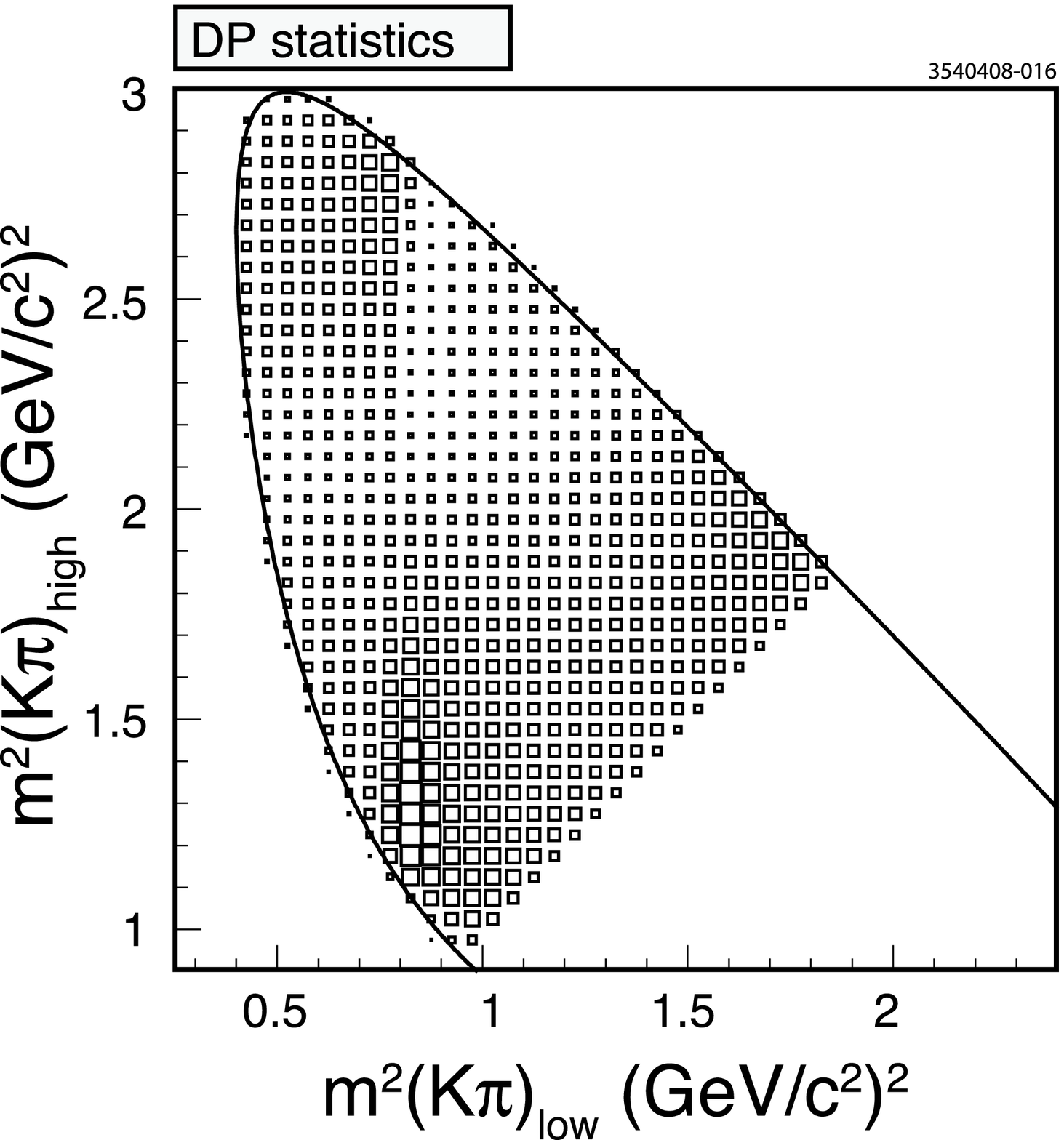}
      \includegraphics[width=0.51\textwidth]{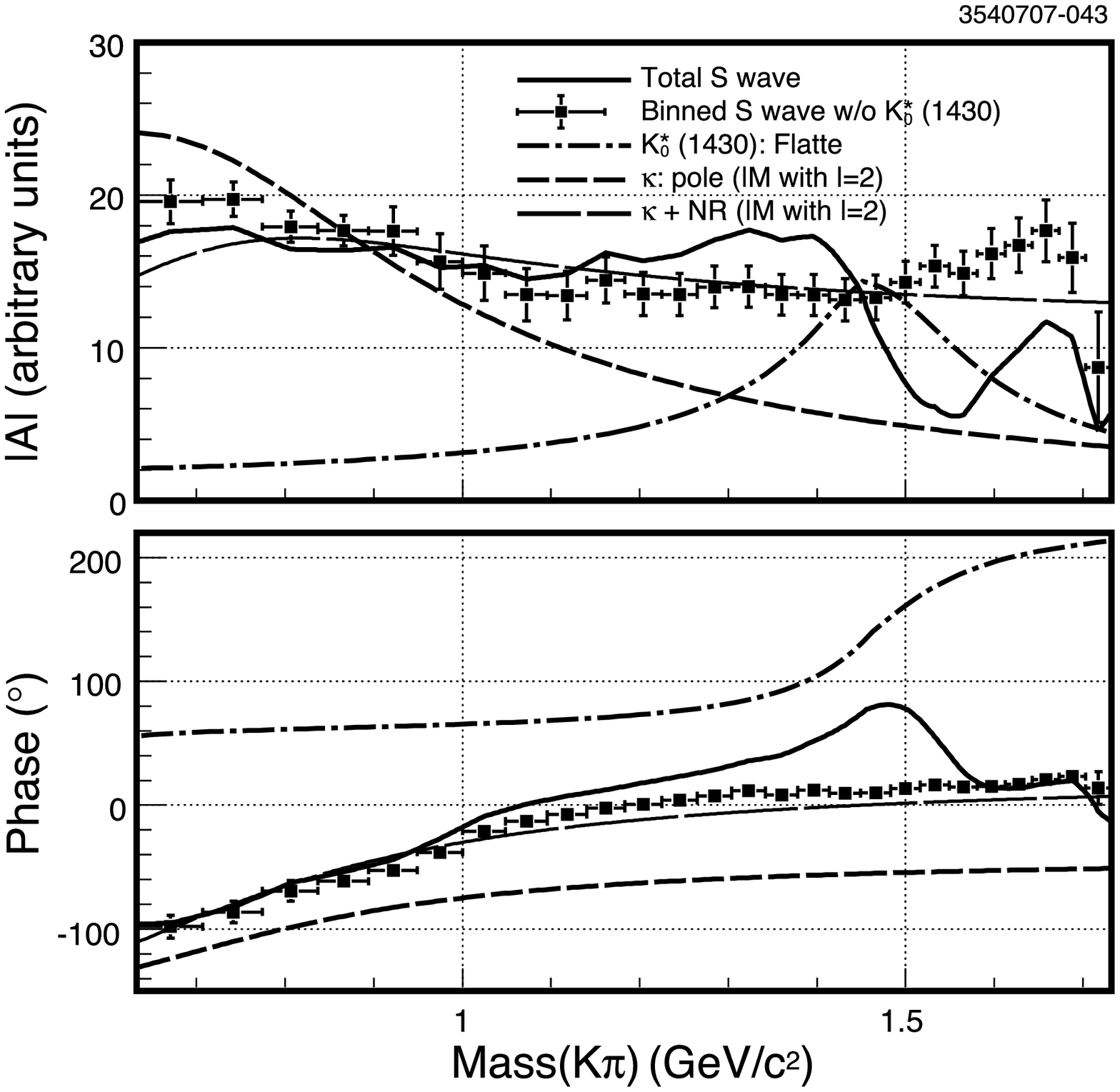}}
\caption{Left: Dalitz plot for $D^+ \to K^- \pi^+ \pi^+$; right:  amplitude
and phase of $I=1/2$ $K \pi$ S wave.  From Ref.\ \cite{Bonvicini:2008jw}.
%\mbox{\includegraphics[height=2.2in]{3540108-008.eps}
%      \includegraphics[height=2.2in]{3540108-012.eps}}
%\caption{Left: Dalitz plot for $D^+ \to K^- \pi^+ \pi^+$ (data); middle:
%corresponding Monte Carlo distribution; right: amplitude and phase of $I=1/2$
%$K \pi$ S wave.  From Ref.\ \cite{Bonvicini:2008jw}.
\label{fig:kmpppp}}
\end{figure}

The CLEO Dalitz plot for $D^+ \to K^- \pi^+ \pi^+$ is shown in Fig.\
\ref{fig:kmpppp}.  The enhancements on the opposite sides of the $K^*(892)$
band for high and low $m^2(K \pi)_{\rm high}$ indicate interference with
an amplitude of opposite parity to $K^*(892)$ (likely an S wave).  Several
fits all have a prominent low-$m(K \pi)$ S-wave amplitude.  A
quasi-model-independent partial wave analysis (QMIPWA; cf.\ \cite{E791})
has a $\kappa$-like behavior, but displaced in overall phase.

% This is Figure 1

In fits to the Dalitz plot, a $\kappa$ pole position depends on whether its
Breit-Wigner width $\Gamma$ is constant or energy-dependent.  The QMIPWA
determines the S-wave $K \pi$ amplitude and phase for 26 $m^2(K \pi)$ bins;
the result resembles the $\kappa$ + nonresonant amplitude obtained in other
fits.  A high-$m(\pi \pi)$ contribution from an $I_{\pi \pi}=2$ amplitude,
perhaps due to $\pi^+ \pi^+ \to \rho^+ \rho^+$, is required for a good fit.
In fits with a $\kappa$, the S-wave phase does not go through $90^\circ$
exactly at the resonance mass.  When $\kappa$ is represented by a complex
pole (equivalent to a constant Breit-Wigner width), the $\kappa \pi^+$ fit
fraction is $\sim 20\%$.

Scalar strange resonances
couple much more strongly to $K \eta'$ than to $K \eta$ \cite{HJLeta}.  In a
limit (corresponding to a particular octet-singlet mixing) where $\eta \simeq
(u \bar u + d \bar d - s \bar s)/\sqrt{3}$; $\eta'\simeq (u \bar u + d \bar d +
2 s \bar s)/\sqrt{6}$, the contributions of strange and nonstrange quarks in
$\eta$ cancel exactly in $K_0^* \to K \eta$ while they add constructively in
$K_0^* \to K \eta'$.  This is the same physics that favors $B \to K \eta'$ over
$B \to K \eta$.  The pattern would be reversed for vector strange resonances.
Thus $K_0^*(1430)$ should be strongly associated with the nearby $K \eta'$
threshold; the $K \pi$ S wave should become inelastic only above this energy.
One then might expect a zero (a manifestation of the Ramsauer-Townsend effect!)
in the scalar $I=1/2$ $K \pi$ amplitude below $K_0^*(1430)$ if a low-mass
$\kappa$ exists.

CLEO and E791 find the $K_0^*(1430)$ heavier ($m \simeq 1460$ MeV) and narrower
($\Gamma \simeq 175$ MeV) than the PDG world average from 2006 \cite{PDG06}
(1414$\pm$6 MeV, 290$\pm$21 MeV, based mainly on elastic $K \pi$ scattering
\cite{LASS}).  BES \cite{Ablikim:2005kp} find a scalar $K \pi$ resonance in
$\chi_{c0} \to \pi^+ \pi^- K^+ K^-$ with $m = (1455 \pm 20 \pm 15)$ MeV,
$\Gamma = 270 \pm 45 ^{+35}_{-30}$ MeV.

% This is Figure 2
\begin{figure}
\includegraphics[width=0.90\textwidth]{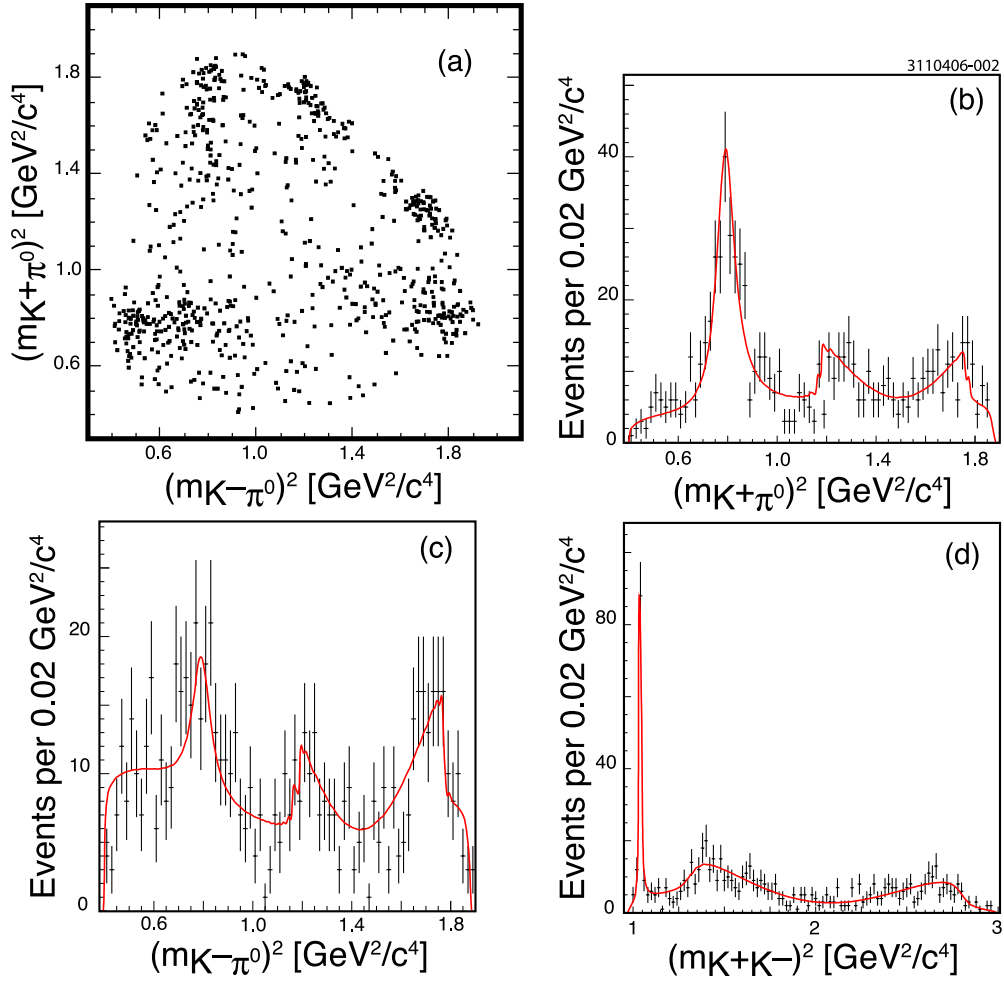}
\caption{(a) Dalitz plot for $D^0 \to K^+ K^- \pi^0$ and projections on (b)
$m^2_{K^+ \pi^0}$; (c) $m^2_{K^- \pi^0}$; (d) $m^2_{K^+ K^-}$.
From Ref.\ \cite{Cawlfield:2006hm}.  Curves denote a fit with $K*(892)$,
$\phi$, and nonresonant S wave.
\label{fig:kpkmpz}}
\end{figure}

\section{$D^0 \to K^+ K^- \pi^0$}

A sample of 735 $D^0 \to K^+ K^- \pi^0$ candidates was obtained with the CLEO
III detector using 9.0 fb$^{-1}$ at 10.58 GeV.  The corresponding Dalitz plot
and its projections are shown in Fig.\ \ref{fig:kpkmpz}.  One sees opposite
signs of the interference between $K^\pm$ and a large S-wave amplitude (typical
fit fraction 20--40\%), implying opposite relative phases for $D^0 \to (K^{*+}
K^-,~ K^{*-} K^+)$.  Although the $K \pi$ S wave is appreciable, one cannot
tell if it is resonant.  A curious dip in $m(K \pi)$ occurs around 1 GeV.
Could this be a Ramsauer-Townsend zero between a $\kappa$ and $K_0^*(1430)$?
The BaBar Collaboration \cite{:2007dc} has $> 11,000$ events in a 385 fb$^{-1}$
sample, with no dip seen.

% This is Figure 3
\begin{figure}
\mbox{\includegraphics[width=0.50\textwidth]{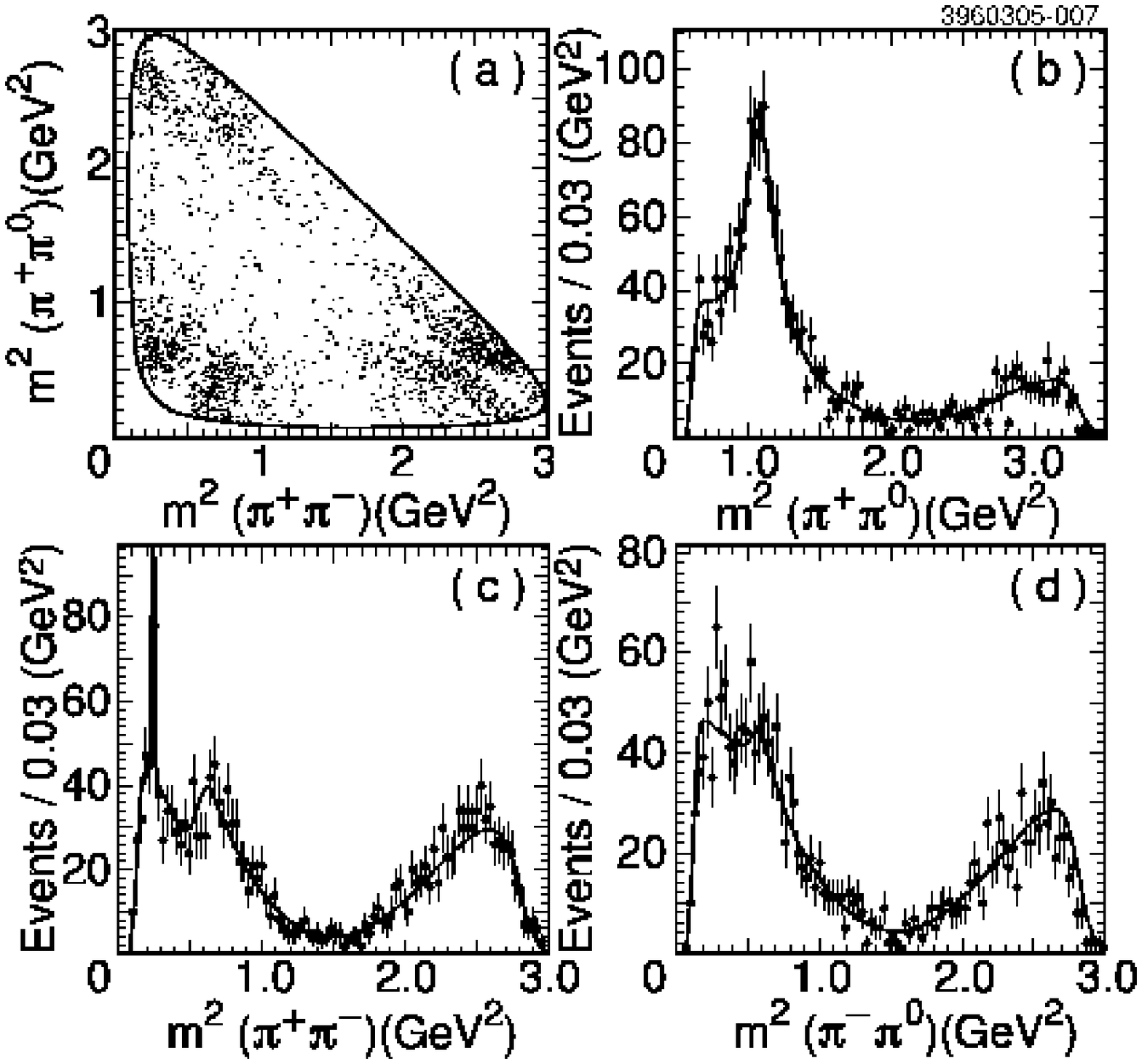} \hskip 0.1in
      \includegraphics[width=0.44\textwidth]{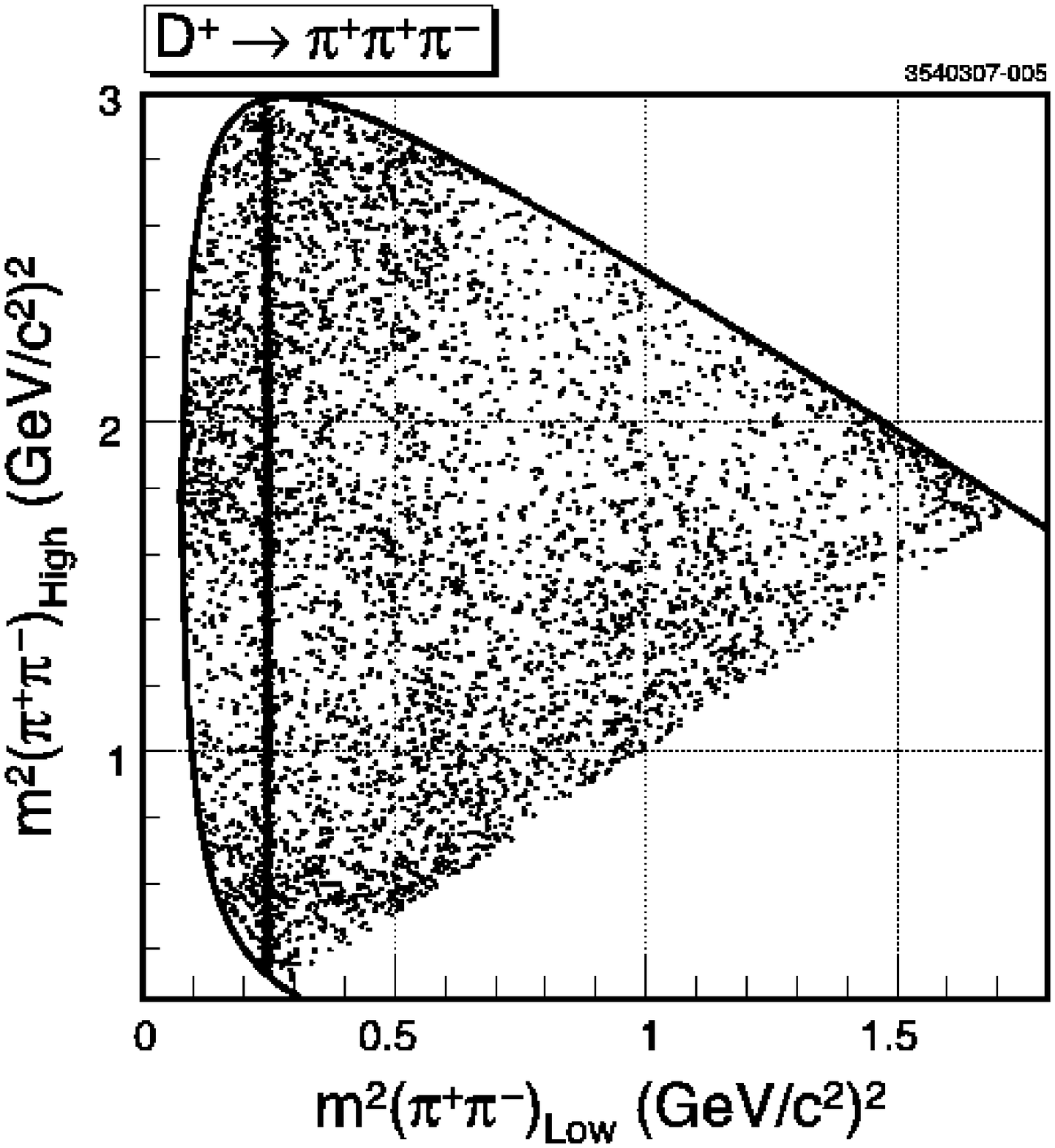}}
\caption{Left: (a) Dalitz plot for $D^0 \to \pi^+ \pi^- \pi^0$; (b)
$m^2_{\pi^+ \pi^0}$ projection; (c) $m^2_{\pi^+ \pi^-}$ projection;
(d) $m^2_{\pi^- \pi^0}$ projection.  Right:  Dalitz plot for $D^+ \to
\pi^- \pi^+ \pi^-$.  The dark vertical band is due to $K_S \to \pi^+ \pi^-$.
\label{fig:ppp}}
\end{figure}

\section{$D^0 \to \pi^+  \pi^- \pi^0$ vs.\ $D^+ \to \pi^- \pi^+ \pi^+$}

CLEO $D^0 \to \pi^+  \pi^- \pi^0$ data are based on 9 fb$^{-1}$ near 10 GeV
\cite{CroninHennessy:2005sy}, while $D^+ \to \pi^- \pi^+ \pi^+$ data are based
on 281 pb$^{-1}$ sample (about 1/3 of the final total) at the $\psi(3770)$
\cite{Bonvicini:2007tc}.  Their Dalitz plots are compared in Fig.\
\ref{fig:ppp}.  While $D^0 \to \pi^+ \pi^- \pi^0$ is dominated by $\rho^\pm,
\rho^0$, $D^+ \to \pi^- \pi^+ \pi^+$ can have only $\rho^0$, not produced by
the charged weak current, so it is not surprising that the scalar fit fraction
is larger in this decay.  For $D^0$ it is found to be $< 4\%$; for $D^+$ it is
40--80\%.

\section{$D^0 \to K_S \pi^0 \pi^0$}

\begin{figure}
\includegraphics[width=0.43\textwidth]{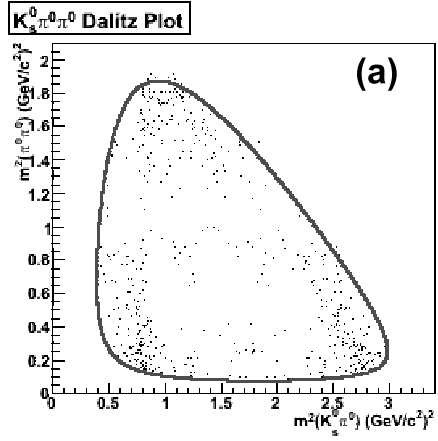}
\caption{Dalitz plot for $D^0 \to K_S \pi^0 \pi^0$.
\label{fig:kzpzpz}}
\end{figure}

A preliminary analysis of $D^0 \to K_S \pi^0 \pi^0$ based on 281 pb$^{-1}$
taken by CLEO at $\psi(3770)$ \cite{Naik:2007es} obtains fit fractions in
the Dalitz plot [Fig.\ \ref{fig:kzpzpz}, showing $m^2(\pi^0\pi^0)$ vs.\
$m^2(K_S \pi^0)$] of (54.2$\pm$5.4$\pm$3.0$\pm$5.0)\% for $K^*(892)$,
(9.0$\pm$3.2$\pm$0.9$\pm$2.7)\% for $f_0(980)$, (23.8$\pm$7.1$\pm$4.7$\pm$8.6)%
\% for $f_0(1370)$, and (11.4$\pm$2.7$\pm$2.1$\pm$3.2)\% for a $K^*(1680)$ with
spin 1.  Judgment on a low-mass $\sigma$ awaits analysis of the full 818
pb$^{-1}$ data sample.

% This is Figure 4

% This is Figure 5
\begin{figure}[!htbp]
\mbox{\includegraphics[width=0.36\textwidth]{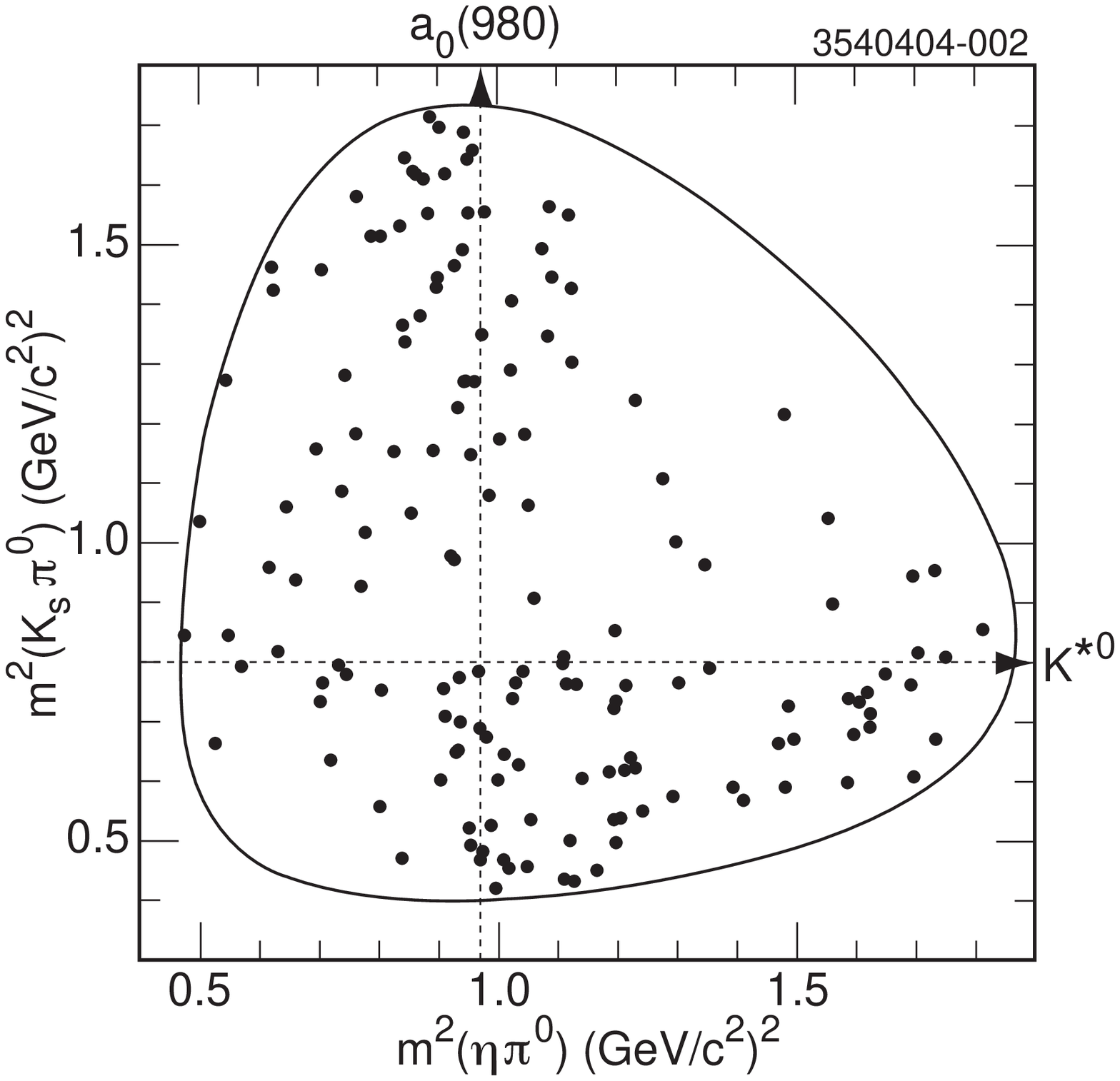}
      \includegraphics[width=0.36\textwidth]{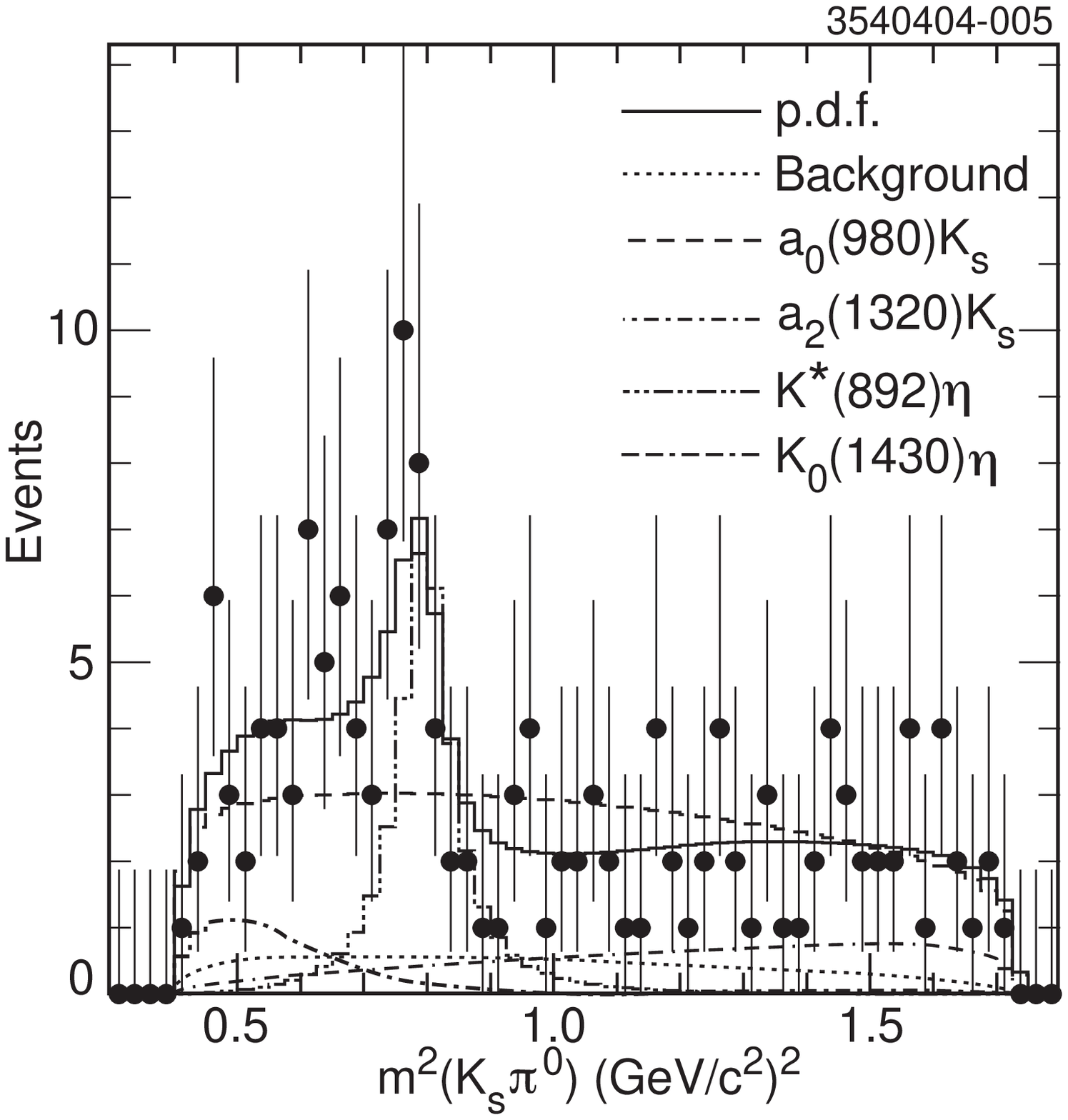}}
\end{figure}
\begin{figure}[!htbp]
\mbox{\includegraphics[width=0.36\textwidth]{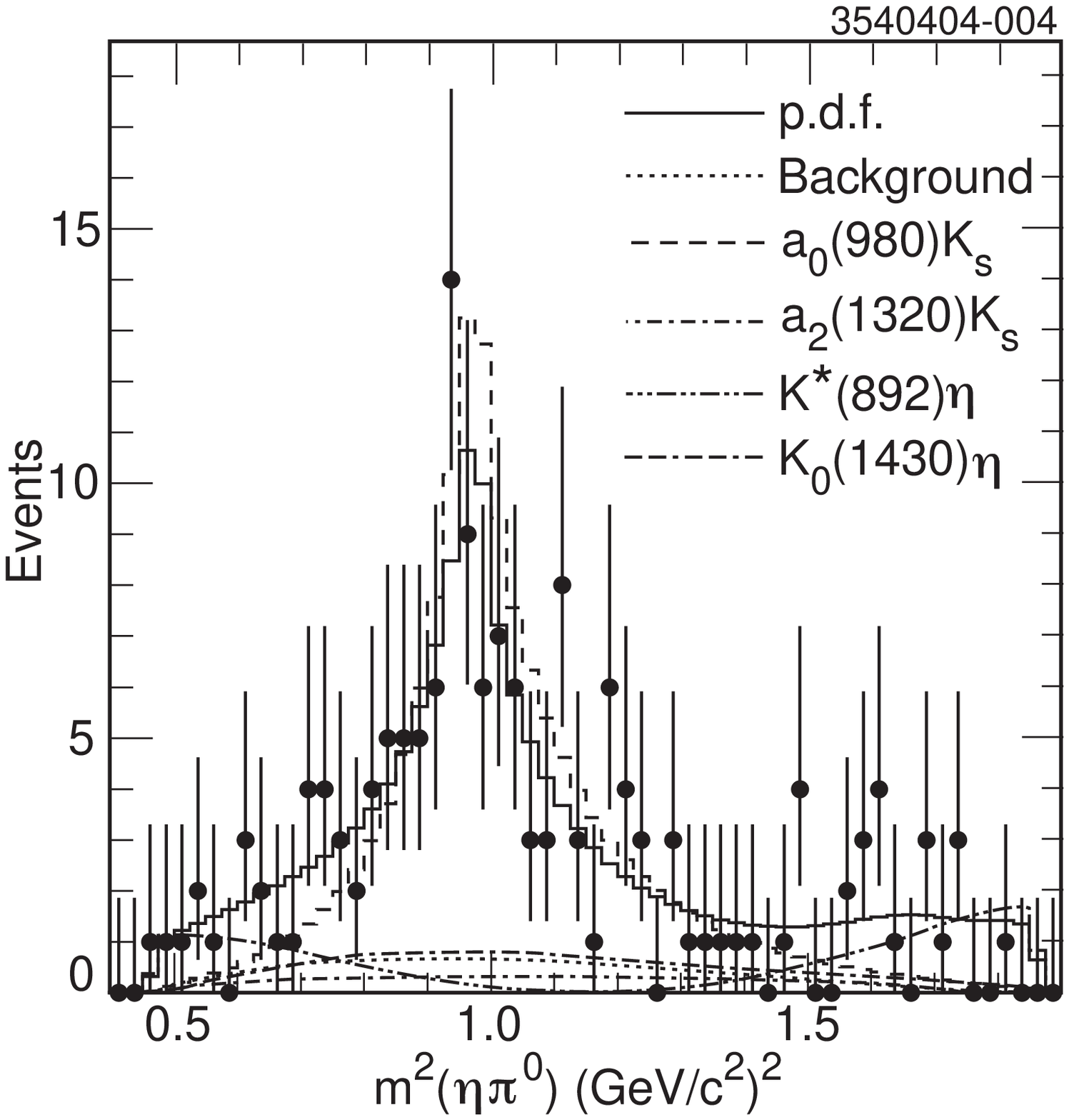}
      \includegraphics[width=0.36\textwidth]{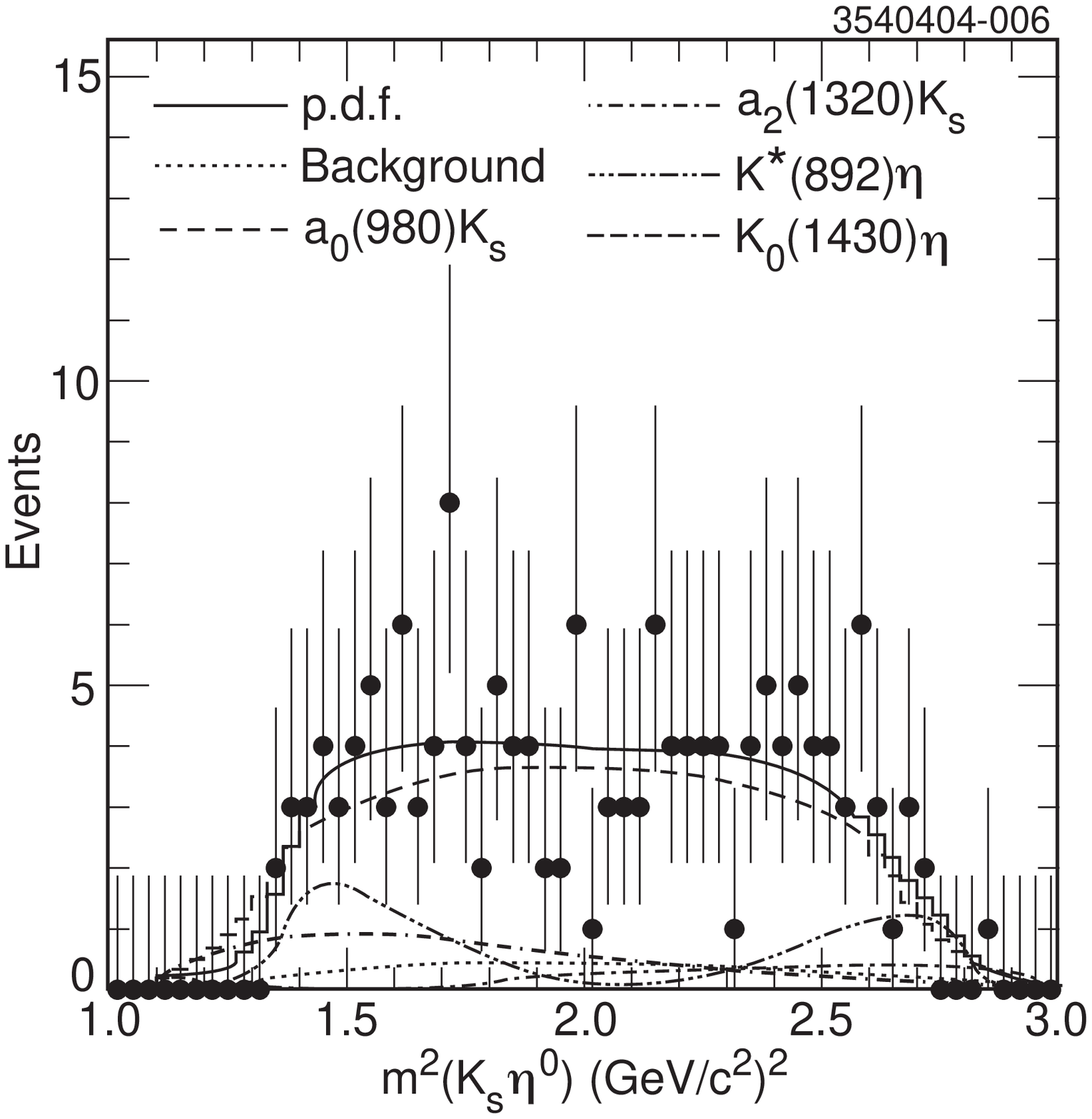}}
\caption{Upper left:  Dalitz plot for $D^0 \to K_S \eta \pi^0$;
upper right:  $m^2_{K_S \pi^0}$ projection; lower left: $m^2_{\eta \pi^0}$
projection; lower right: $m^2_{K_S \eta}$ projection.
\label{fig:kzetapz}}
\end{figure}
\newpage

\section{$D^0 \to K_S \eta \pi^0$}

Published CLEO data on $D^0 \to K_S \eta \pi^0$ come from 9.0 fb$^{-1}$ near
the $\Upsilon(4S)$, yielding a signal of $155 \pm 22$ events.  The Dalitz
plot (Fig.\ \ref{fig:kzetapz}) is dominated by $a_0(980) K_S$ [fit fraction
${\cal O}(1)$].  The $K^*(892) \eta$ fit fraction is $\simeq 30\%$.  It would
be interesting to compare $D^0 \to K_S a_0^0$ with $D^0 \to K^- a_0^+$ and
$D^+ \to K_S a_0^+$.  Related processes are $D^0 \to \bar \kappa^0 \pi^0$, $D^0
\to \kappa^- \pi^+$, and $D^+ \to \bar \kappa^0 \pi^+$ if $\kappa$ and $a_0$
belong to the same SU(3) multiplet.  The $\psi(3770)$ data set will contain
$\sim 1200~K_S \pi^0 \eta$, $\sim 8000~K^- \pi^+ \eta$, and $\sim 5000~K_S
\pi^+ \eta$.

\section{Three-body $\chi_c$ decays}

\begin{figure}
\mbox{\includegraphics[width=0.33\textwidth]{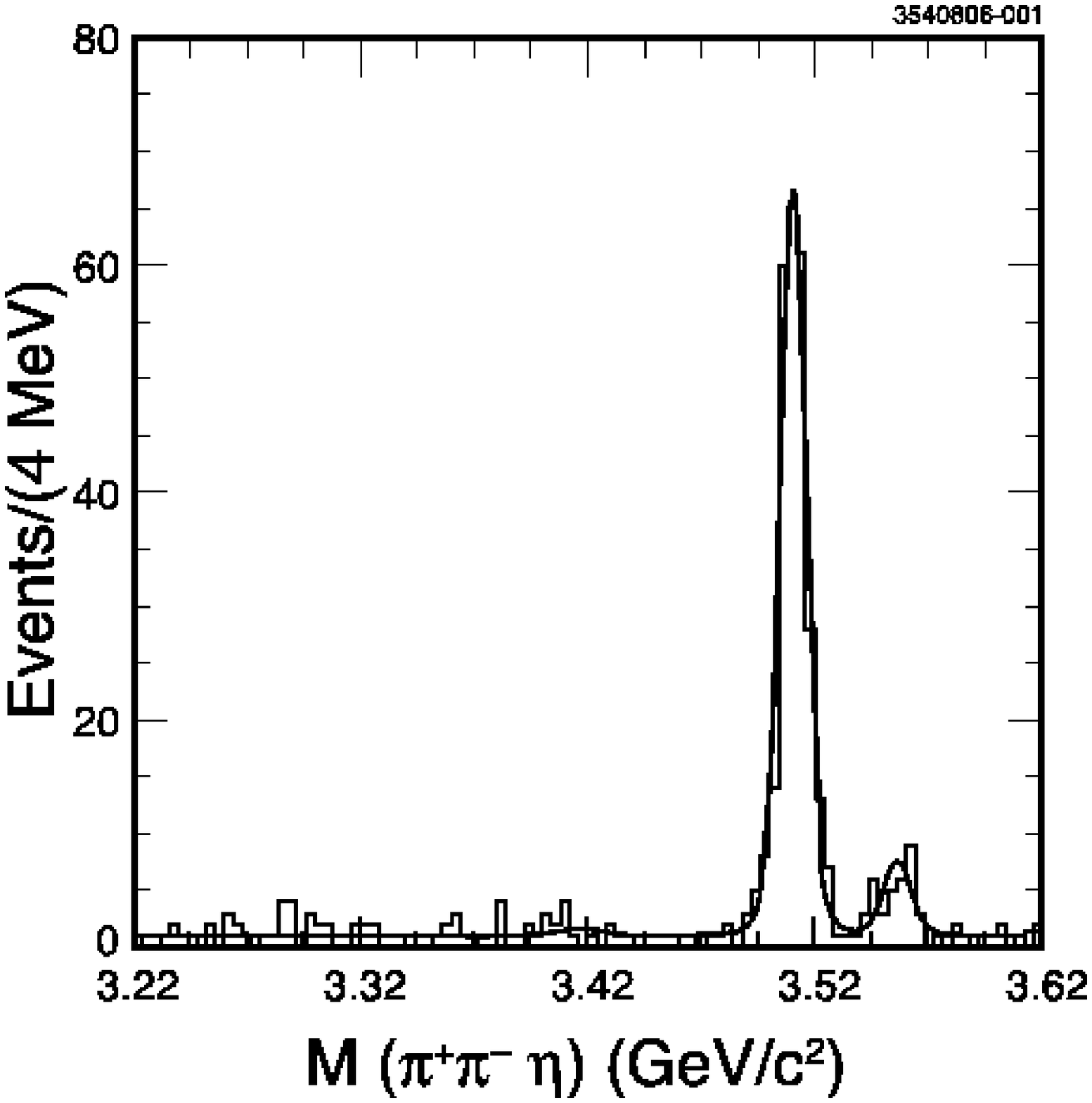}
      \includegraphics[width=0.33\textwidth]{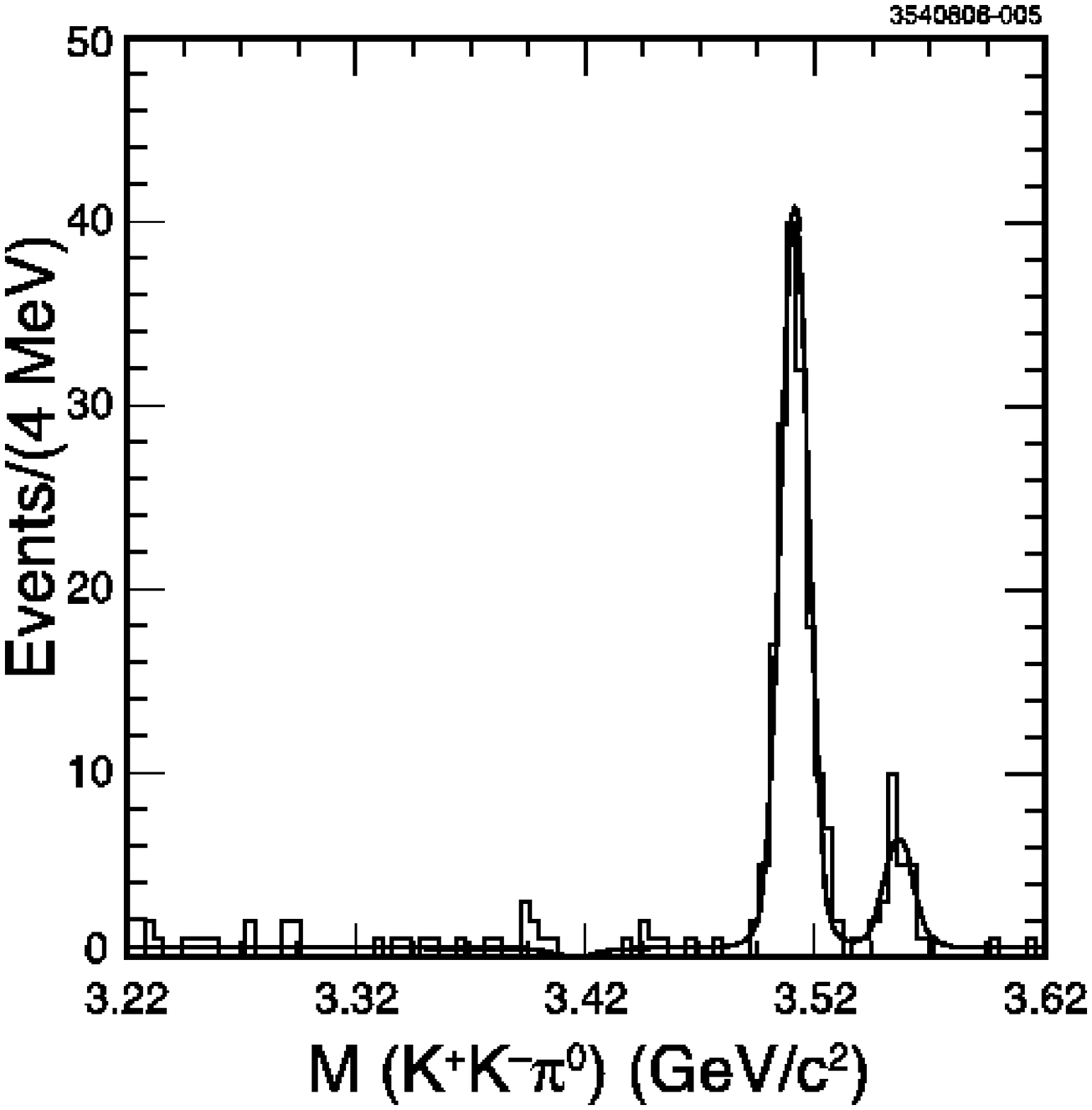}
      \includegraphics[width=0.33\textwidth]{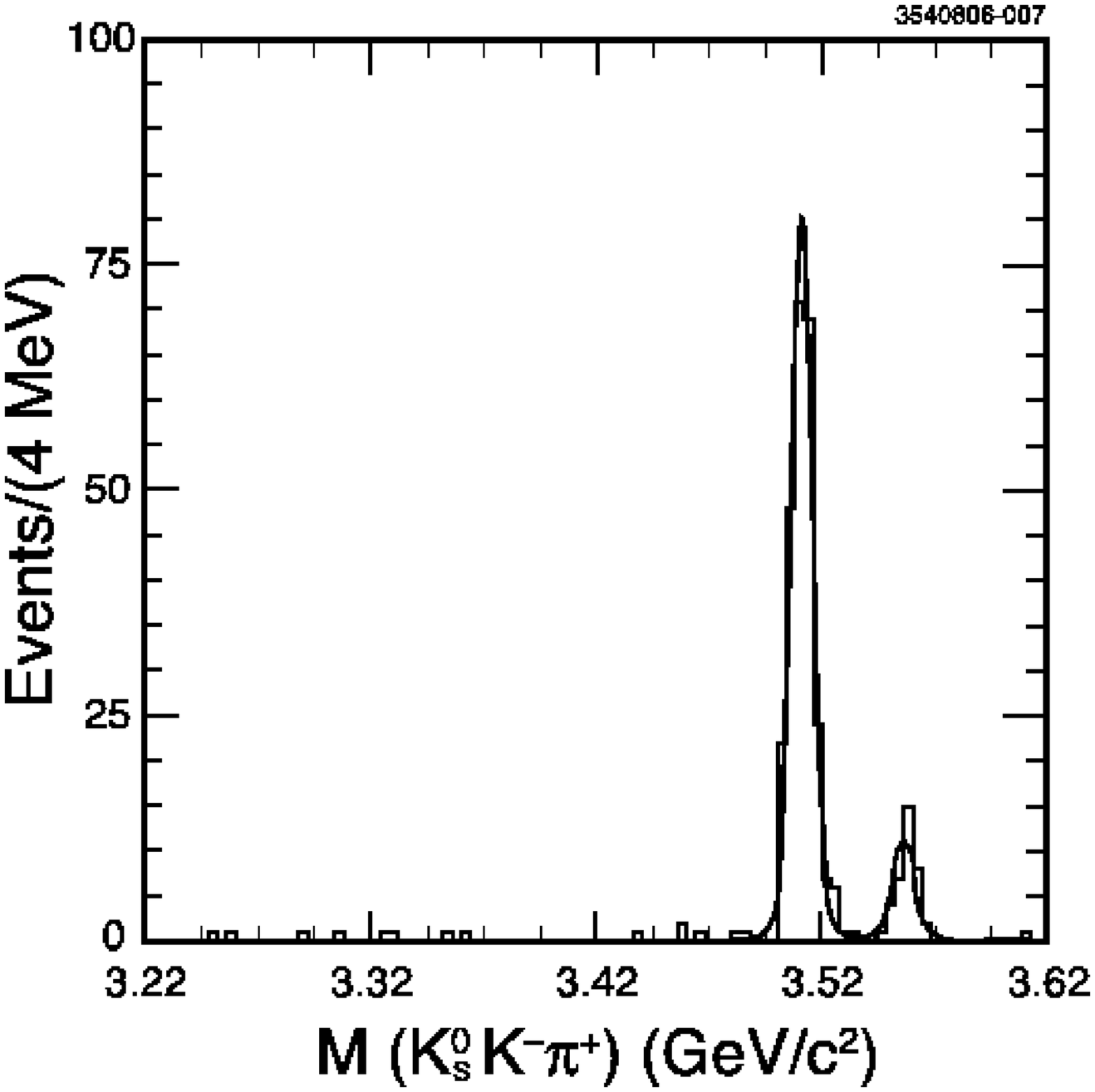}}
\caption{Mass distributions for final states $X$ in $\psi(2S) \to \gamma X$.
Left: $X = \pi^+ \pi^- \eta$; middle: $X = K^+ K^- \pi^0$; right: $X = 
K_S K^- \pi^+$.
\label{fig:chic}}
\end{figure}

The transitions $\psi(2S) \to \gamma \chi_{cJ}$ ($J=0,1,2$) were studied
by CLEO, reconstructing exclusive final states for 3 million $\psi(2S)$
\cite{Athar:2006gh} (24.5 million more $\psi(2S)$ are under analysis).  The
signals are shown in Fig.\ \ref{fig:chic}.  The three-body decays $\chi_{c1}
\to (\eta \pi^+ \pi^-,~K^+ K^- \pi^0,~K_S K^\pm \pi^\mp)$ have enough events
($255^{+17}_{-16}$, $157 \pm 13$, and $249 \pm 16$, respectively) for Dalitz
plot analyses.  In these channels $I_{\pi \pi} = 0$ in $\chi_{c1} \to \eta
\pi^+ \pi^-$ and $I_{K \bar K} = 1$ in $\chi_{c1} \to (K^+ K^- \pi^0,~K_S K^\pm
\pi^\mp)$.  The analysis of the 3 million $\psi(2S)$ did not consider
$\chi_{c1}$ polarization or interference between resonances, desirable features
in analysis of the full sample.

% This is Figure 6

\subsection{$\chi_{c1} \to \eta \pi^+ \pi^-$}

\begin{figure}
\includegraphics[width=0.96\textwidth]{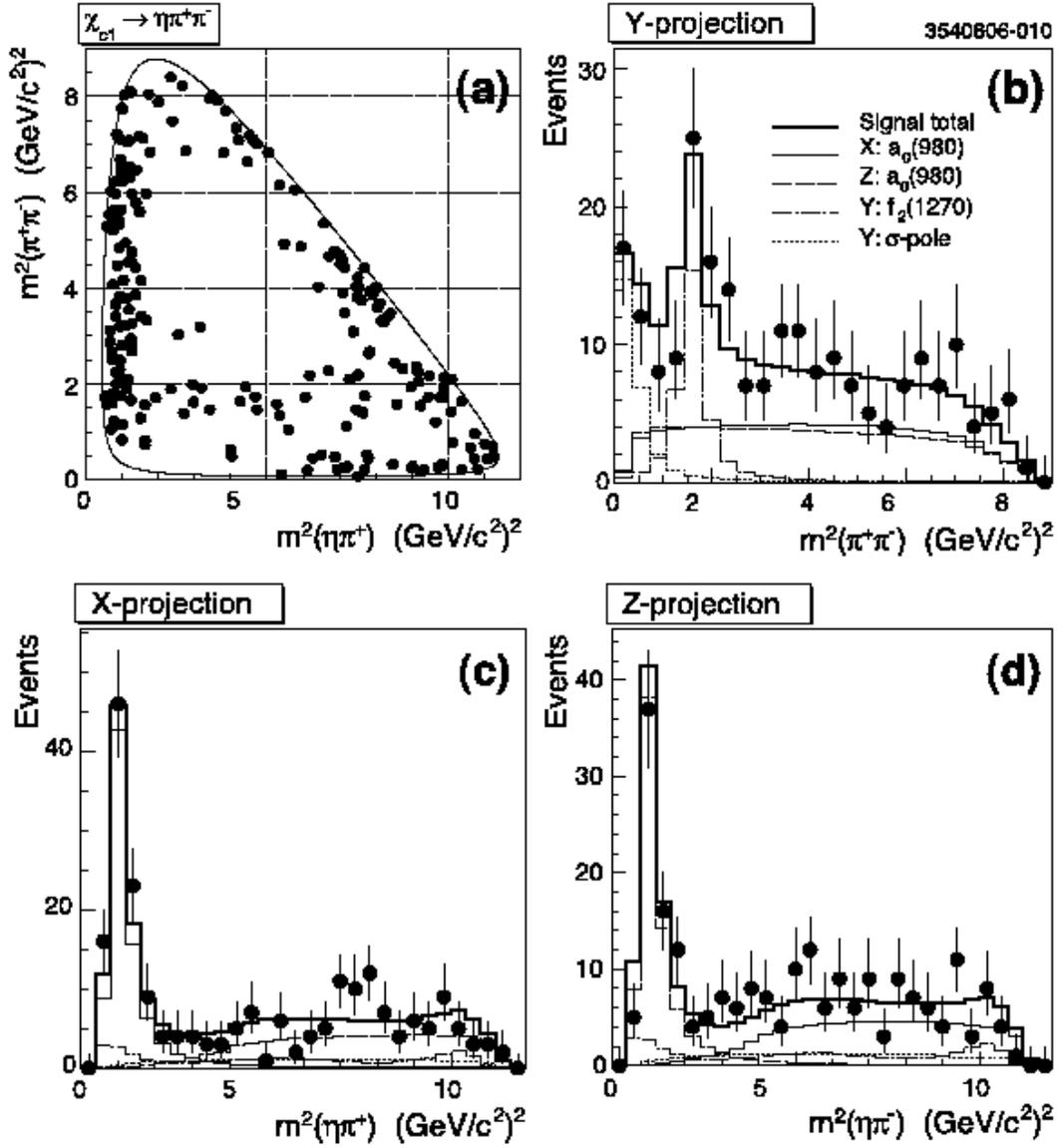}
\caption{(a) Dalitz plot for $\chi_{c1} \to \eta \pi^+ \pi^-$ and projections
on (b) $m^2_{\pi^+ \pi^-}$; (c) $m^2_{\eta \pi^+}$; (d) $m^2_{\eta \pi^-}$.
Main contributions to $m^2_{\pi^+ \pi^-}$ projection are $\sigma$ and
$f_2(1270)$; main contributions to $m^2_{\eta \pi^\pm}$ projections are
$a_0(980)^\pm$.
\label{fig:etapp}}
\end{figure}

The Dalitz plot for $\chi_{c1} \to \eta \pi^+ \pi^-$ and its projections are
shown in Fig.\ \ref{fig:etapp}.  The fit fractions, in percent, are
75.1$\pm$3.5$\pm$4.3 for $a_0(980)^\pm \pi^\mp$, 14.4$\pm$3.1$\pm$1.9 for
$f_2(1270)\eta$, and 10.5$\pm$2.4$\pm$1.2 for $\sigma \eta$.  Here $\sigma$ is
parametrized by a complex pole at (511$\pm$28--$i$102$\pm$50) MeV.  The
low-mass $\pi \pi$ enhancement is visible both in the Dalitz plot and in the
$m^2(\pi^+ \pi^-)$ projection.  Flavor SU(3) would imply that $\chi_{c1} \to
\kappa \bar K$ should be visible if $\chi_{c1} \to a_0 \pi$ is so prominent.

% This is Figure 7

\subsection{$\chi_{c1} \to (K^+ K^- \pi^0,~K_S K^\pm \pi^\mp)$}

\begin{figure}
\includegraphics[width=0.96\textwidth]{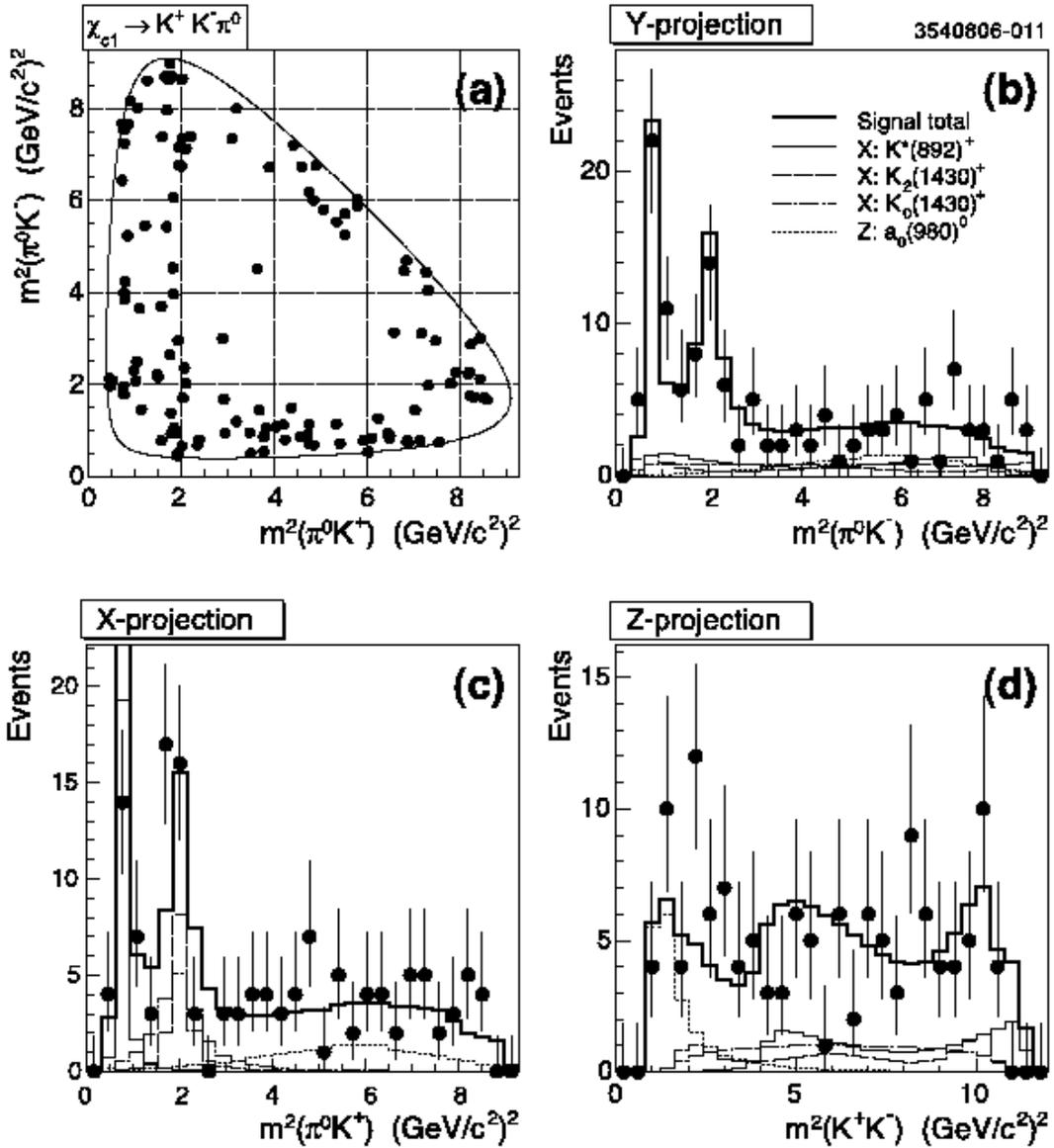}
\caption{(a) Dalitz plot for $\chi_{c1} \to K^+ K^- \pi^0$ and projections on
(b) $m^2_{\pi^0 K^-}$; (c) $m^2_{\pi^0 K^+}$; (d) $m^2_{K^+ K^-}$.  Main
contributions to $m^2_{\pi^0 K^\pm}$ projections are $K^*(892)$, $K_2(1430)$,
and $K_0(1430)$; main contribution to $m^2_{K^+ K^-}$ projection is
$a_0(980)^0$.
\label{fig:pzkpkm}}
\end{figure}

\begin{figure}
\includegraphics[width=0.96\textwidth]{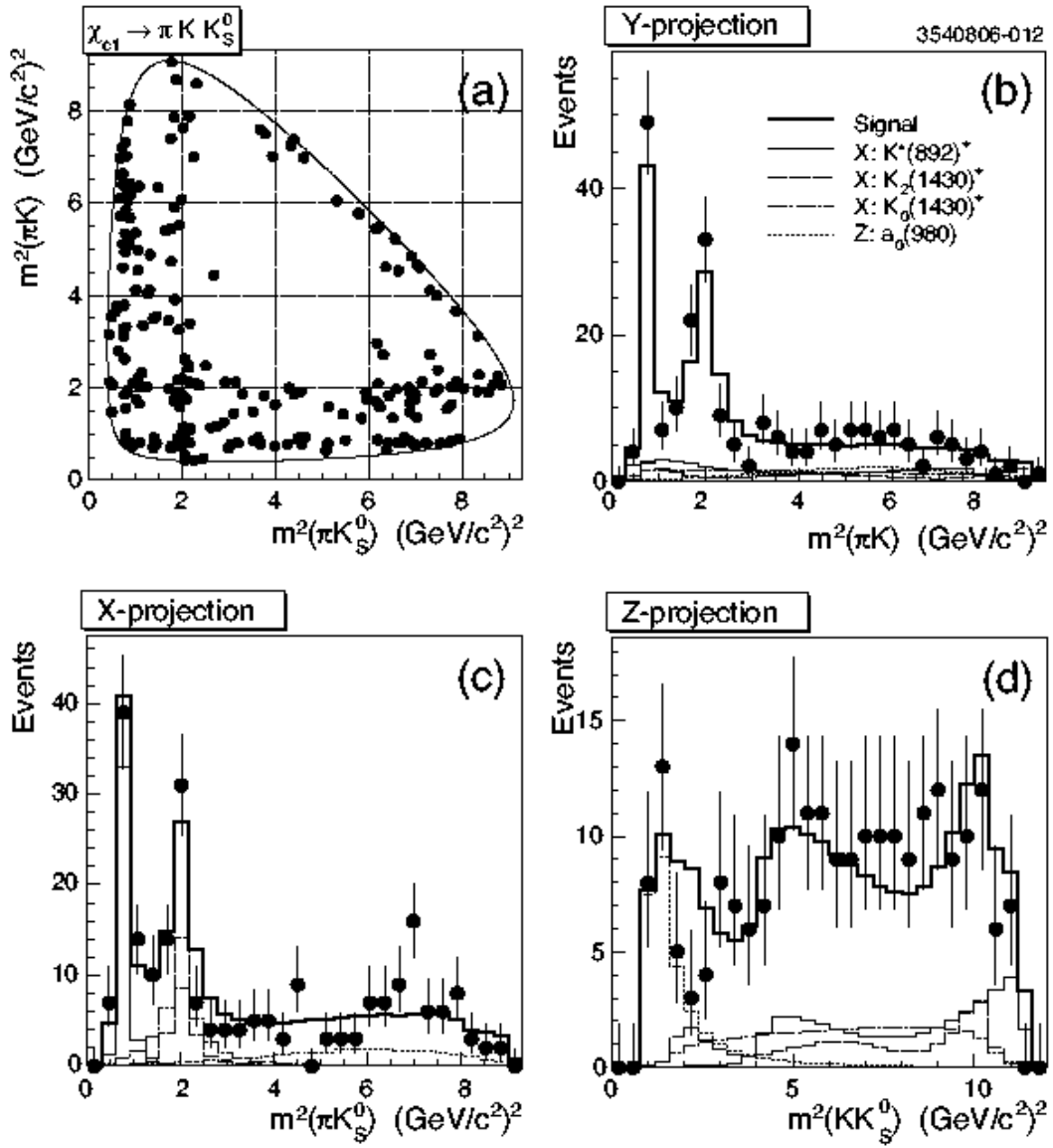}
\caption{(a) Dalitz plot for $\chi_{c1} \to K_S K^\pm \pi^\mp$ and projections
on (b) $m^2_{\pi^\pm K^\mp}$; (c) $m^2_{\pi^\pm K_S}$; (d) $m^2_{K^\pm K_S}$.
Main contributions are as in Fig.\ \ref{fig:pzkpkm}.
\label{fig:kspk}}
\end{figure}

The Dalitz plots for $\chi_{c1} \to (K^+ K^- \pi^0,~K_S K^\pm \pi^\mp)$ and
their projections are shown in Figs.\ \ref{fig:pzkpkm} and \ref{fig:kspk}.
In analyzing them the $I=0$ channels of $K^+ K^-\pi^0$ and $K_S K^\pm \pi^\mp$,
related by isospin, have been combined.  The fit fractions, in percent, are
31.2$\pm$2.2$\pm$1.7 for $K^*(892) \bar K$, 30.4$\pm$3.5$\pm$3.7 for
$K_0^*(1430) \bar K$, 23.1$\pm$3.4$\pm$7.1 for $K_2^*(1430) \bar K$, and
15.1$\pm$2.7$\pm$1.5 for $a_0(980) \pi$.  The addition of a $\kappa$ or
nonresonant $K \pi$ S-wave doesn't improve the fit quality.  However, account
of interference might show a low-mass $K \pi$ S wave as in the analysis of $D^+
\to K^- \pi^+ e^+ \nu_e$ \cite{Shepherd:2006tw}.

% This is Figure 8

% This is Figure 9

For $\chi_{c1} \to K_S K^\pm \pi^\mp$ one expects twice as many events as
in $\chi_{c1} \to K^+ K^- \pi^0$, neglecting efficiency differences.  In the
full data set the expected sample of $\sim 2000~\chi_{c1} \to \eta \pi^+ \pi^-$
should permit a good determination of the ``$\sigma$'' properties.  We expect
many other three-body $\chi_{cJ}$ final states to be reconstructed in the full
CLEO $\psi(2S)$ radiative decay sample.

\section{Conclusions}

Charmed meson and charmonium decays can be a rich source of information on
scalar resonances between pairs of pseudoscalar mesons.  The data from CLEO
show the potential of these channels.  The $a_0(980)$ and $f_0(980)$ are
without question in CLEO data; $\kappa(800)$ and $\sigma(600)$ make sporadic
appearances.  Their inferred masses and widths depend on production channels
and line shape models.  Because of Bose statistics, the $\sigma~(I_{\pi \pi} =
0)$ is easier to isolate than the $\kappa~(I_{K \pi} = 1/2)$.  Although not
related to charm decays, it is notable that CLEO needs a $\sigma$ in describing
$\tau \to \pi \pi^0 \pi^0 \nu$ \cite{Asner:1999kj}.  Tests of whether the
$a_0(980)$ and $f_0(980)$ belong to a nonet with $\kappa(800)$ and
$\sigma(600)$ are available in charmonium (e.g., $\chi_{c1}$) decays.  We look
forward to the realization of CLEO's ultimate potential for shedding light on
scalar mesons with the analysis of the full data sets from $\psi(2S)$,
$\psi(3770)$, and $E_{\rm cm} = 4170$ MeV.

\begin{theacknowledgments}
 I thank my colleagues on CLEO and particularly Mikhail Dubrovin for much
valuable advice, and the organizers of this Workshop for providing a
stimulating environment for discussion.  This work was supported in part
by the United States Department of Energy under Grant No.\ DE FG02 90ER40560. 
\end{theacknowledgments}


\begin{thebibliography}{9}

\bibitem{:2007nn}
  G.~Bonvicini {\it et al.} [CLEO Collaboration], CLEO-CONF 07-01,
  %``Dalitz plot analysis of the D+ -> K- pi+ pi+ decay,''
  arXiv:0707.3060 [hep-ex].
  %%CITATION = ARXIV:0707.3060;%%

\bibitem{Muramatsu:2002jp}
  H.~Muramatsu {\it et al.} [CLEO Collaboration],
  %``Dalitz analysis of D0 --> K0(S) pi+ pi-. ((B)),''
  \emph{Phys.\ Rev.\ Lett.\ }{\bf 89}, 251802 (2002)
  [Erratum-\emph{ibid.\ }{\bf 90}, 059901 (2003)].
  %[arXiv:hep-ex/0207067].
  %%CITATION = PRLTA,89,251802;%%

\bibitem{Pappagallo:2007zz}
  M.~Pappagallo [BaBar Collaboration], SLAC-PUB-12983, published in
  %``Charm Dalitz analyses at BaBar,''
  \emph{Acta Phys.\ Polon.\ B} {\bf 38}, 2885 (2007).
  %%CITATION = APPOA,B38,2885;%%

\bibitem{Poluektov:2006ia}
  A.~Poluektov {\it et al.} [Belle Collaboration],
  %``Measurement of phi(3) with Dalitz plot analysis of B+ --> D(*) K(*)+
  %decay,''
  \emph{Phys.\ Rev.\ D} {\bf 73}, 112009 (2006).
  %[arXiv:hep-ex/0604054].
  %%CITATION = PHRVA,D73,112009;%%

\bibitem{CroninHennessy:2005sy}
  D.~Cronin-Hennessy {\it et al.} [CLEO Collaboration],
  %``Searches for CP violation and pi pi S-wave in the Dalitz-plot of D0 -->
  %pi+ pi- pi0,''
  \emph{Phys.\ Rev.\ D} {\bf 72}, 031102 (2005)
  [Erratum-\emph{ibid.\ D}{\bf 75}, 119904 (2007)].
  %[arXiv:hep-ex/0503052].
  %%CITATION = PHRVA,D72,031102;%%

\bibitem{Bonvicini:2007tc}
  G.~Bonvicini {\it et al.} [CLEO Collaboration],
  %``Dalitz plot analysis of the D+ --> pi- pi+ pi+ decay,''
  \emph{Phys.\ Rev.\ D} {\bf 76}, 012001 (2007).
  %[arXiv:0704.3954 [hep-ex]].
  %%CITATION = PHRVA,D76,012001;%%

\bibitem{Cawlfield:2006hm}
  C.~Cawlfield {\it et al.} [CLEO Collaboration],
  %``Measurement of interfering K*+ K- and K*- K+ amplitudes in the decay D0
  %--> K+ K- pi0,''
  \emph{Phys.\ Rev.\ D} {\bf 74}, 031108 (2006).
  %[arXiv:hep-ex/0606045].
  %%CITATION = PHRVA,D74,031108;%%

\bibitem{Rubin:2004cq}
  P.~Rubin {\it et al.} [CLEO Collaboration],
  %``First observation and Dalitz analysis of the D0 --> K0(S) eta pi0 decay,''
  \emph{Phys.\ Rev.\ Lett.\ }{\bf 93}, 111801 (2004).
  %[arXiv:hep-ex/0405011].
  %%CITATION = PRLTA,93,111801;%%

\bibitem{LASS} D. Aston {\it et al.} [LASS Collaboration], \emph{Nucl.\
Phys.\ }{\bf B296}, 496 (1988).

\bibitem{Naik:2007es}
  P.~Naik, L.~Zhang and N.~Lowrey [CLEO Collaboration], presented at HADRON 07,
  %``Dalitz Plot Analyses at CLEO-c,''
  arXiv:0712.2266.
  %%CITATION = ARXIV:0712.2266;%%

\bibitem{Athar:2006gh}
  S.~B.~Athar {\it et al.} [CLEO Collaboration],
  %``chi/cJ decays to h+ h- h0,''
  \emph{Phys.\ Rev.\ D} {\bf 75}, 032002 (2007).
  %[arXiv:hep-ex/0607072].
  %%CITATION = PHRVA,D75,032002;%%

\bibitem{Bonvicini:2008jw}
  G.~Bonvicini {\it et al.} [CLEO Collaboration],
  %``Dalitz plot analysis of the D^+ -> K^-pi^+pi^+ decay,''
  arXiv:0802.4214 [hep-ex].
  %%CITATION = ARXIV:0802.4214;%%

\bibitem{E791} E. M. Aitala {\it et al.} [Fermilab E791 Collaboration],
  \emph{Phys.\ Rev.\ Lett.\ }{\bf 89}, 121801 (2002);
  \emph{Phys.\ Rev.\ D} {\bf 73}, 032004 (2006);
  \emph{ibid.\ }{\bf 74}, 059901(E) (2006).

\bibitem{HJLeta} H. J. Lipkin, \emph{Phys.\ Rev.\ Lett.\ }{\bf 46}, 1307
(1981); \emph{Phys.\ Lett.\ B} {\bf 254}, 247 (1991).

\bibitem{PDG06} W.-M. Yao {\it et al.} [Particle Data Group], \emph{J.\
Phys.\ G} {\bf 33}, 1 (2006).

\bibitem{Ablikim:2005kp}
  M.~Ablikim {\it et al.} [BES Collaboration],
  %``Partial wave analysis of chi/c0 --> pi+ pi- K+ K-,''
  \emph{Phys.\ Rev.\ D} {\bf 72}, 092002 (2005).
  %[arXiv:hep-ex/0508050].
  %%CITATION = PHRVA,D72,092002;%%

\bibitem{:2007dc}
  B.~Aubert {\it et al.} [BABAR Collaboration],
  %``Amplitude Analysis of the decay D^0 --> K^- K^+ pi^0,''
  \emph{Phys.\ Rev.\ D} {\bf 76}, 011102 (2007).
  %[arXiv:0704.3593 [hep-ex]].
  %%CITATION = PHRVA,D76,011102;%%

\bibitem{Shepherd:2006tw}
  M.~R.~Shepherd {\it et al.} [CLEO Collaboration],
  %``Model independent measurement of form factors in the decay D+ --> K-  pi+
  %e+ nu/e,''
  \emph{Phys.\ Rev.\ D} {\bf 74}, 052001 (2006).
  %[arXiv:hep-ex/0606010].
  %%CITATION = PHRVA,D74,052001;%%

\bibitem{Asner:1999kj}
  D.~M.~Asner {\it et al.} [CLEO Collaboration],
  %``Hadronic structure in the decay tau- --> nu/tau pi- pi0 pi0 and the  sign
  %of the tau neutrino helicity,''
  \emph{Phys.\ Rev.\ D} {\bf 61}, 012002 (2000).
  %[arXiv:hep-ex/9902022].
  %%CITATION = PHRVA,D61,012002;%%
\end{thebibliography}
\end{document}